\documentclass{article}

\usepackage[utf8]{inputenc}
\usepackage{tismir}
\usepackage{amsmath}
\usepackage{hyperref}
\usepackage{url}
\usepackage{graphicx}
\usepackage{booktabs}
\usepackage{lipsum}
\usepackage{array}
\usepackage[table]{xcolor}  

\title{A Lightweight Two-Branch Architecture for Multi-Instrument Transcription via Note-Level Contrastive Clustering}

\author{
Ruigang Li\thanks{Southeast University, No. 2 Southeast University Road, Nanjing 211189, Jiangsu, China} ~and Yongxu Zhu\protect\footnotemark[1]
}

\date{}

\type{research}

\authorref{Ruigang Li, and Yongxu Zhu}
\journalyear{2026}
\journalvolume{9}
\journalissue{1}
\journalpages{119--130}
\doi{10.5334/tismir.300}

\authorshort{Ruigang Li et al}
\titleshort{Lightweight Transcription via Note-Level Clustering}

\begin{document}


\twocolumn[{%
\maketitleblock
\begin{abstract}
Existing multi-timbre transcription models struggle with generalization beyond pre-trained instruments, rigid source-count constraints, and high computational demands that hinder deployment on low-resource devices. We address these limitations with a lightweight model that extends a timbre-agnostic transcription backbone with a dedicated timbre encoder and performs deep clustering at the note level, enabling joint transcription and dynamic separation of arbitrary instruments given a specified number of instrument classes. Practical optimizations including spectral normalization, dilated convolutions, and contrastive clustering further improve efficiency and robustness. Despite its small size and fast inference, the model achieves competitive performance with heavier baselines in terms of transcription accuracy and separation quality, and shows promising generalization ability, making it highly suitable for real-world deployment in practical and resource-constrained settings.
\end{abstract}

\begin{keywords}
automatic music transcription, instrument separation, deep clustering, low-resource
\end{keywords}
}]
\saythanks{}


\section{Introduction}\label{sec:introduction}
Automatic Music Transcription (AMT), which converts audio signals into symbolic musical notation, represents a fundamental challenge in Music Information Retrieval (MIR) \citep{Overview}. While current systems achieve remarkable accuracy in transcribing polyphonic music for specific instrument timbres \citep{OnsetsAndFrames,Guitar,violinTrans}, we investigate a more practical scenario: processing audio mixtures containing $K$ instrument classes and outputting $K$ tracks of musical notes, each corresponding to a distinct timbre. We term this task ``timbre-separated transcription'', which differs from conventional source separation by operating at the symbolic level to extract note representations rather than reconstructing audio waveforms. In this paper, we study the musical sounds produced by instruments, and the term ``timbre'' refers specifically to instrument categories (e.g., violin, viola, and flute), distinguished from individual physical sources.

Current solutions often use unified classification-based models that treat timbre identification as a categorization task \citep{attentionAMT, MT3}, which limits their flexibility. They demand extensive training data, fix the maximum number of separable sources, and fail to generalize to unseen timbres, essentially functioning as a ``timbre dictionary''. Moreover, these models are typically large and computationally demanding, making them inaccessible to general users despite advances in transcription accuracy.

To address these issues, we propose a lightweight two-branch architecture that decouples pitch/onset estimation from timbre representation learning and employs deep clustering \citep{DeepClustering} to achieve timbre-separated transcription. Our key advance over prior deep clustering transcription \citep{DeepSphericalClustering} is note-level clustering (vs. frame-level), tailored for symbolic output. The first branch performs timbre-agnostic transcription, predicting frame-level activations with a compact, fully convolutional network. The second branch learns direction-aware timbre embeddings, which are clustered at the note level to enable dynamic instrument separation. Furthermore, we discuss the impact of dataset and possible improvements. All code is open-sourced, with the model already deployed in a web-based assistive transcription tool\endnote{Link to the web tool: \url{https://madderscientist.github.io/noteDigger/}}.

\section{Background and Related Work}\label{sec:bg}
\subsection{AMT Fundamentals}
The majority of existing AMT approaches employ deep neural networks to transform audio inputs into piano-roll-like representations, characterized as two-dimensional matrices indexed by time frames and note pitches. The process generally comprises: 1) time–frequency feature extraction, 2) frame-level probability estimation with neural networks, and 3) note generation through binarization or specialized output networks.

Stage 1 is typically implemented with classical signal-processing transforms. The Short-Time Fourier Transform (STFT) yields linear-spaced frequency bins, whereas the Mel-scale Transform (Mel) and Constant-Q Transform (CQT) provide logarithmic spacing that better matches musical pitch perception. However, these interpretable transforms may not always be optimal, leading to explorations of learnable encodings \citep{TasNet,sincnet,LEAF} and hybrid methods \citep{CombineSpectralTemporal,Demucs}. The CQT is readily enriched into a Harmonic CQT (HCQT) \citep{HCQT_origin} by vertically shifting the spectrogram according to the bin offsets of successive harmonics and concatenating the shifted copies along an additional axis \citep{HCQT}. This prior-driven expansion allows small convolutional kernels to focus efficiently on musically relevant frequency components.

Stage 2 produces a two-dimensional posteriogram that encodes the probability of pitch activation at each time-frequency bin. Convolutional layers leverage the spectro-temporal grid to capture local structures \citep{semanticSegmentation,BasicPitch}, while temporal-sequence models extend the receptive field along time, modeling long-range dependencies \citep{OnsetsAndFrames}. To translate frame-wise posteriors into discrete notes, in stage 3, most systems rely on onset detection (i.e., the initial frame of each note) and then link successive active frames within an onset-defined segment to form complete note events \citep{OnsetsAndFrames,attentionAMT,TransformerAgnostic,BasicPitch}.

\subsection{Timbre-Separated Transcription}
Multi-instrument timbre-separated transcription goes beyond traditional AMT by extracting instrument-specific notes from polyphonic mixtures. \citet{TAPE} demonstrated that simply cascading source separation followed by transcription leads to suboptimal results due to error propagation. Therefore, it is necessary to combine transcription and separation in a mutually reinforcing manner.

Innovative methods have emerged for this challenge. \citet{MT3} introduced a sequence-to-sequence AMT model that outputs multiple instrument tracks with explicit instrument assignments, achieving high accuracy but relying heavily on diverse multi-timbre data. Multi-task learning frameworks like Timbre-Trap \citep{TimbreTrap} and Cerberus \citep{Cerberus} show that joint optimization can lead to synergistic effects. \citet{attentionAMT} proposed a self-attention-based instance segmentation approach that performs joint note detection and instrument classification, achieving timbre-separated transcription for a closed set of trained instruments.

For handling unseen timbres, researchers have explored different strategies. The zero-shot learning method proposed by \citet{ZeroShot} employs a query-based mechanism: under contrastive supervision, it learns to encode a clean reference example into a timbre embedding that generalizes to unseen instruments. This embedding is then used to sequentially modulate the U-Net layers, enabling zero-shot separation and transcription of audio sources with novel timbres not encountered during training. The work by \citet{DeepSphericalClustering} pioneered the application of deep clustering networks to timbre-separated transcription. Their approach develops a generalizable timbre encoder and enables separation with a non-fixed number of classes through clustering, showing how these tasks can complement each other when properly combined.

\subsection{Deep Clustering Methodology}
Deep clustering serves as a fundamental approach addressing two core challenges in source separation: the permutation problem and speaker-independent separation. The methodology operates by encoding each time-frequency bin into a high-dimensional feature, then assigning labels through clustering algorithms to generate separation masks.

The pioneering work of \citet{DeepClustering} first applied deep clustering to separate unknown speakers. This was later extended by \citet{DANet}, who introduced learnable attractors as speaker-specific reference points in an embedding space that cluster time-frequency bins, enabling end-to-end separation of a variable number of sources. The adaptation to music transcription was first achieved by \citet{DeepSphericalClustering}, but their method uses transcription mainly to aid audio separation, lacks dedicated note-creation postprocessing, and omits key implementation details. More fundamentally, these studies share two critical limitations: clustering at the frame level produces fragmented results, and the hard-assignment strategy cannot handle overlapping situations. In contrast, our clustering postprocess is tailored for music transcription, aggregating frame embeddings into coherent note events with minimal computational overhead.

\section{Proposed Method}

\begin{figure*}[t]
  \centering
  \includegraphics[width=\linewidth]{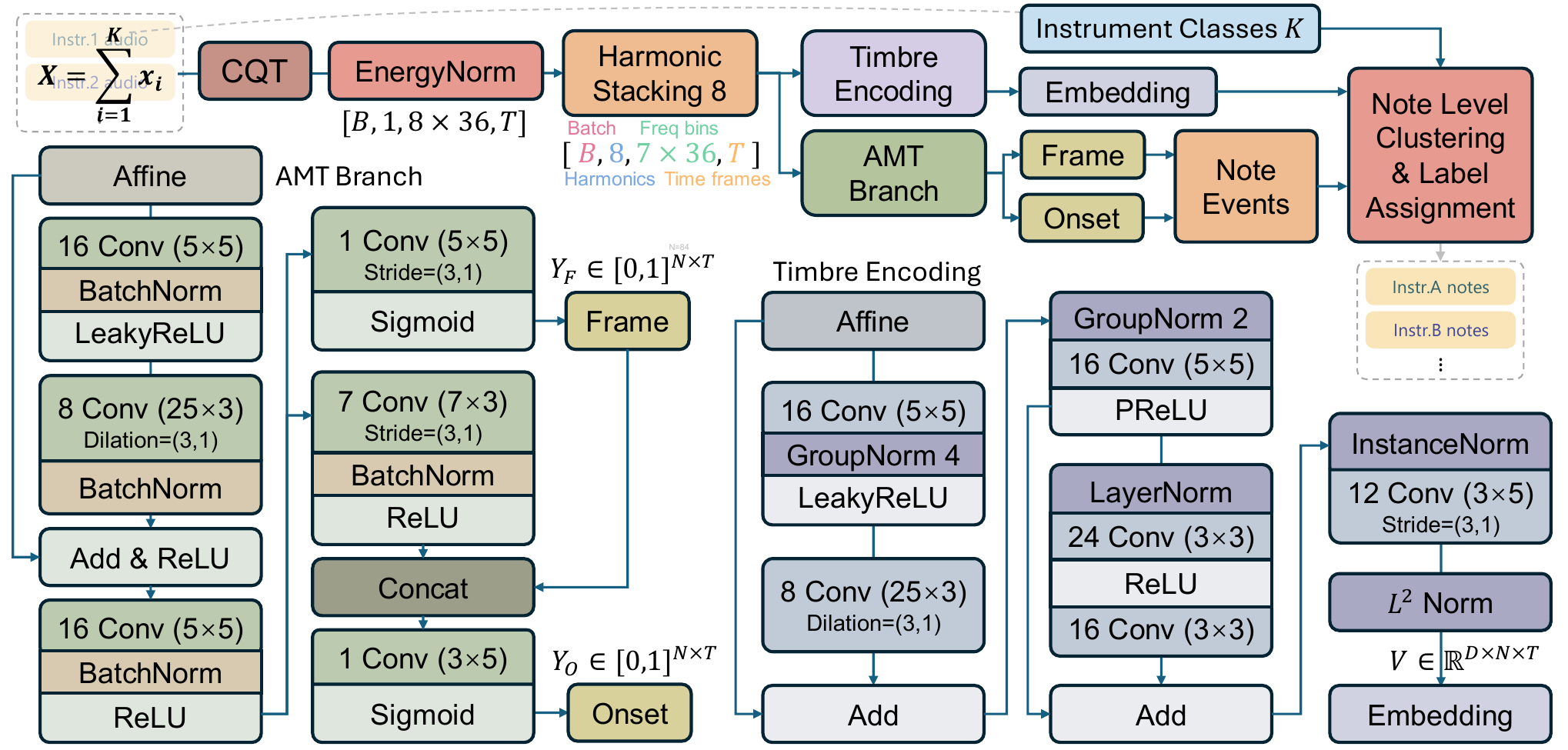}
  \caption{Overview of the proposed method. \textbf{Top:} Overall pipeline, which takes a multi-timbral mixture audio as input and outputs note events for each constituent timbre. \textbf{Bottom left:} AMT branch producing timbre-agnostic transcription outputs—frame activation posteriorgram $Y_F$ and onset activation posteriorgram $Y_O$. \textbf{Bottom right:} Timbre encoding branch yielding a $D$-dimensional timbre embedding $V$ for each time–frequency bin, where $N=84$ denotes the target pitch range.}
\label{fig:architecture}
\end{figure*}

\subsection{Problem Configuration}
Let $ X = \sum_{i=1}^K x_i \in \mathbb{R}^l $ be the mixed audio signal of $K$ instrument classes (timbres), where $ l = 22.05 \text{ [kHz]} \times \text{Time} \text{ [sec]} $ is the signal length, and $ x_i $ is the single timbre audio corresponding to the frame activation $ y_{Fi} \in [0,1]^{N \times T} $ and onset activation $ y_{Oi} \in [0,1]^{N \times T} $. Since we are aiming for the pitch range $ C_1 \sim B_7 $, we set $ N = 7 \times 12 = 84 $. The goal of timbre-agnostic transcription is to map $ X $ to $ \sum_{i=1}^K y_{Fi} $ and $ \sum_{i=1}^K y_{Oi} $, while the goal of timbre-separated transcription is to obtain $ \{ y_{Fi} \}_{i=1}^K $. First, we apply the CQT to obtain $ Q \in \mathbb{C}^{F \times T} $, where $ T $ is the number of frames and $ F $ is the number of frequency points analyzed. We extended the CQT analysis to eight octaves with three bins per semitone, ensuring each note has at least two harmonics and yielding $ F = 8 \times 12 \times 3 = 288 $ bins.
Our proposed model outputs the timbre-agnostic transcription results $ Y_F = \sum_{i=1}^K \hat{y}_{Fi} $ and $ Y_O = \sum_{i=1}^K \hat{y}_{Oi} $, and the timbre-sensitive representation $ V \in \mathbb{R}^{D \times N \times T} $. Finally, $V$ is clustered to assign labels to $ Y_F $, separating $ \{ \hat{y}_{F1}, \hat{y}_{F2}, \dots, \hat{y}_{FK} \} $ from $ Y_F $.

\subsection{Overall Architecture}
As shown in Figure~\ref{fig:architecture}, our timbre-separated transcription model comprises two parallel branches: timbre-agnostic transcription and timbre encoding, both taking the HCQT as input. We also design a postprocessing pipeline tailored to this task.

The timbre-agnostic transcription branch builds upon \texttt{BasicPitch}~\citep{BasicPitch}, a lightweight fully convolutional network that demonstrates strong performance using only local contextual information. We identify several limitations and introduce targeted optimizations, yielding a more compact architecture with comparable accuracy.

The timbre encoding branch adopts a similar fully convolutional design. To incorporate global context, we employ InstanceNorm and GroupNorm for adaptive normalization across the entire input. While convolutional architectures are more deployment-friendly on low-resource devices than RNN-based approaches commonly used in deep clustering \citep{DeepClustering, DeepSphericalClustering}, they may lack long-range modeling capacity; alternative designs are discussed in the experiments.

\begin{figure*}[t]
  \centering
  \includegraphics[width=2\columnwidth]{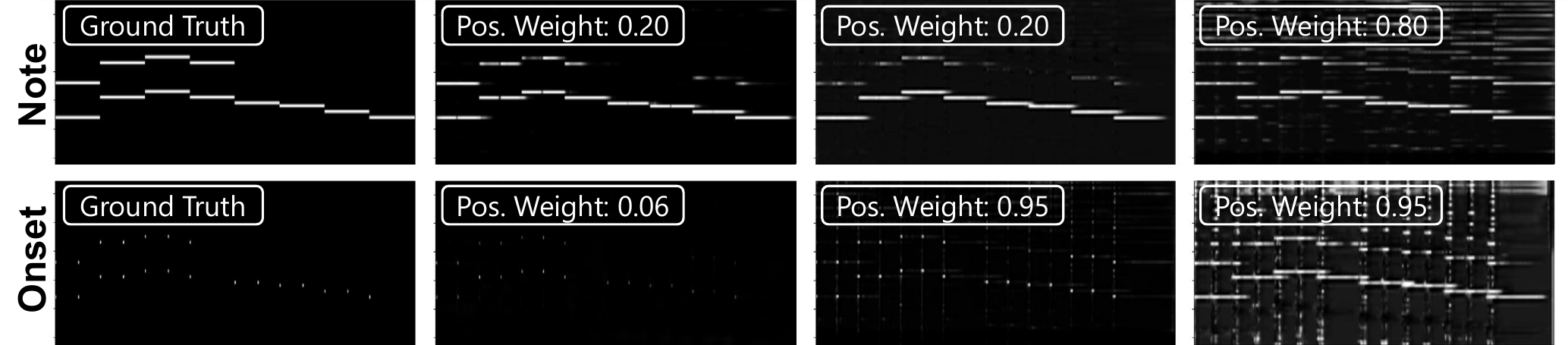}
  \caption{Training outcomes using Focal Loss with varying positive class weights. Each column represents a training session from initialization to convergence.}
\label{fig:loss}
\end{figure*}

\subsection{EnergyNorm for Spectral Normalization}
Conventional two-dimensional normalization schemes often involve subtractive operations and compute statistics directly over the time–frequency plane. These practices alter the relative amplitude relationships among harmonic components within the same analysis frame, thereby undermining the acoustic basis of timbre perception.

We therefore introduce an interpretable normalization approach based on frame‑wise energy. Leveraging Parseval's theorem and assuming a Gaussian time-domain signal, we standardize the sample variance of frame energy to unity, yielding a chi-square distributed amplitude envelope. Given a CQT spectrum $Q \in \mathbb{C}^{F \times T}$, the process is described as:
\begin{align}\label{eq:cqtnorm}
E_t = \sum_{f=1}^{F} | q_{f,t} |^2 &\in \mathbb{R}^{T}, \\
\sigma = \sqrt{Var(E_t)} &\in \mathbb{R}, \\
E_{\text{norm}} = \frac{| Q |^2}{\sigma} &\in \mathbb{R}^{F \times T}.
\end{align}

This approach reduces the normalization complexity from two dimensions to one, significantly lowering computational costs. Furthermore, to address the limitations of strictly non-negative features in neural network learning, we apply a logarithmic transformation followed by an affine transformation to enhance representational capacity. The final normalized representation is defined as:
\begin{equation}
\tilde{Q} = k \cdot (\log | Q |^2 - \log\sigma) + b,
\end{equation}

where $k$ and $b$ are learnable scalars.

\begin{figure*}[t]
  \centering
  \includegraphics[width=1.95\columnwidth]{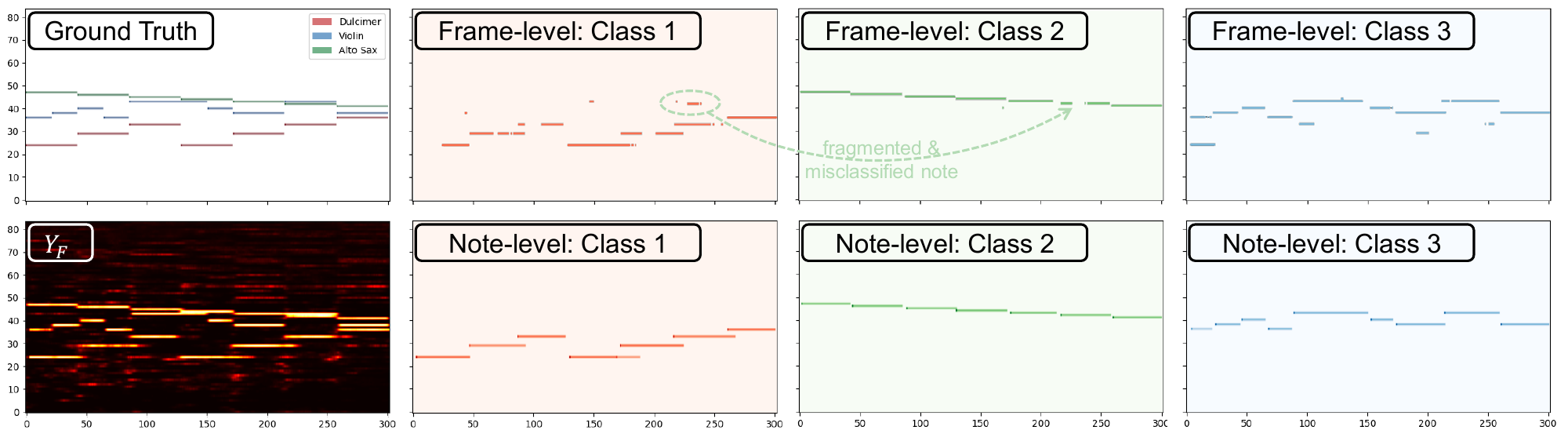}
  \caption{Results of frame-level and note-level postprocessing for triple separation.}
\label{fig:postprocess}
\end{figure*}

\subsection{Dilated Convolution for Harmonic Context}
In \texttt{BasicPitch}, a large convolutional kernel spanning 39 frequency bins (one octave plus one note) is explicitly mentioned as effective for mitigating octave errors. However, the reference implementation applies padding to both ends, effectively limiting the receptive field to half an octave above and below, rather than the intended full range above or below. Given the explicit spatial correspondence between input and output representations, this limitation prevents the model from directly attending to information at harmonic intervals. To address this issue, we adopted a more robust strategy by extending the receptive field to cover two octaves in both directions. We implemented this using a dilated convolution with a frequency-axis kernel size of 25 and a dilation factor of 3. This configuration ensures comprehensive harmonic coverage while simultaneously reducing the parameter count.

\subsection{Focal Loss for Class Imbalance}
\texttt{BasicPitch} employs a class-balanced cross-entropy loss to address the sparsity of onsets, with negative and positive class weights set at 0.05 and 0.95, respectively. However, our experiments revealed that this weighting scheme exacerbates class imbalance. Figure \ref{fig:loss} presents our controlled experimental results, demonstrating that assigning a larger weight to the positive class of onsets leads to severe false positives, indicating that sparse classes should be assigned smaller weights.

We directly used Focal Loss \citep{focalloss} in Eqn.~(\ref{eqn:focalloss}) with $\gamma = 1$, setting the positive class weights for notes and onsets at 0.2 and 0.06, respectively.
\begin{align}\label{eqn:focalloss}
    \begin{array}{c}
        \text{FL}(p_t) = -\alpha_t (1 - p_t)^{\gamma} \log(p_t),\\
        p_t =
        \begin{cases}
            p & \text{if } y = 1 \\
            1 - p & \text{if } y = 0,
        \end{cases}
        \qquad
        \alpha_t =
        \begin{cases}
            \alpha & \text{if } y = 1 \\
            1 - \alpha & \text{if } y = 0.
        \end{cases}
    \end{array}
\end{align}

\subsection{InfoNCE for Contrastive Cluster Formation}
\citet{DeepClustering} and \citet{DeepSphericalClustering} use the following deep clustering objective:
\begin{align}\label{eq:affinityLoss}
\mathcal{L}_{\text{affinity}} &= \|UU^T - ZZ^T\|
_F^2 \nonumber \\
&= \|U^TU\| _F^2 - 2\| U^TZ\| _F^2 + \| Z^TZ\|
_F^2,
\end{align}

where $Z \in \mathbb{R}^{M \times K}$ denotes the one-hot class labels for active bins ($K$: number of classes, $M$: number of bins with ground-truth activation), and $ U^T \in \mathbb{R}^{D \times M} $ is the timbre encodings at the corresponding locations extracted from $ V \in \mathbb{R}^{D \times N \times T}$. The $m$-th column $\mathbf{u_m} \in \mathbb{R}^D$ of $U^T$ represents the timbre encoding of the $m$-th active time-frequency bin. Defining $\hat{Z} = ZZ^T \in \mathbb{R}^{M \times M}$ yields:
\begin{equation}
\hat{z}_{i,j} = \begin{cases}
1& \text{if } \mathbf{u_i} \text{ and } \mathbf{u_j} \text{ should share class}\\
0& \text{otherwise.}
\end{cases}
\end{equation}

This loss function benefits from low spatial complexity and high computational efficiency. However, it is at least three orders of magnitude larger than the AMT loss (frame\&onset transcription with Focal Loss) and converges slowly, which makes joint optimization more difficult when a backbone is shared. Furthermore, this objective enforces orthogonality between encodings of different classes. We argue that strict orthogonality is an overly stringent constraint; for clustering objectives, anti-parallel alignment yields larger inter-class distances than orthogonality. Consequently, we adopt the contrastive InfoNCE loss \citep{InfoNCE} to facilitate cluster formation:

\begin{equation}\label{eqn:infonceloss}
\mathcal{L}_{\text{InfoNCE}} = 
-\sum_{n=1}^{M} \log \frac{
    \displaystyle \sum_{\substack{m=1, m \neq n \\ y_n = y_m}}^{M} 
    \exp\!\left( \frac{\mathbf{u}_n^\top \mathbf{u}_m}{\tau} \right)
}{
    \displaystyle \sum_{\substack{m=1, m \neq n}}^{M} 
    \exp\!\left( \frac{\mathbf{u}_n^\top \mathbf{u}_m}{\tau} \right)
},
\end{equation}

where $y_m$ denotes the ground-truth class label of $\mathbf{u}_m$. Empirically setting the temperature parameter $\tau = 0.15$ aligns the magnitude of the clustering loss with the AMT loss and synchronizes their convergence rates. Crucially, in scenarios involving a single timbre, the InfoNCE loss naturally vanishes. In contrast, the MSE-based loss continues to force feature vectors toward specific directions even in the absence of contrasting classes, causing the mean output to deviate significantly from the origin and resulting in potential instability.

\subsection{Postprocessing for Timbre-Separated Transcription}\label{sec:postprocess}

Existing deep clustering–based source separation methods are predominantly designed for spectrogram reconstruction. In the field of music transcription, however, there is a lack of discussion on methods for creating specific notes from clustering results.

\citet{DeepSphericalClustering} proposed a frame-level transcription method that predicts a binary mask by assigning an additional cluster to "silence bins" and performing K-means clustering over the entire time–frequency space. However, this approach suffers from several critical limitations:

\begin{itemize}
    \item The absence of onset information hinders accurate note creation from masks.
    \item Frame-level separation is prone to fragmenting notes into scattered pieces.
    \item The large number of bins makes clustering algorithms very slow.
    \item Most clustering algorithms require specifying the number of clusters, and the hard-assignment strategy cannot handle overlapping situations, where instruments play the same pitch simultaneously.
\end{itemize}

Benefiting from the dual-branch architecture, the separation process can be conducted at the note level. We first obtain notes using \texttt{BasicPitch}'s note decoding method, then weight and sum the encodings of each frame in the note to get its timbre encoding, which is subsequently clustered via spectral clustering. This method aggregates time-frequency bins of the same note, greatly reducing the number of samples for clustering and mitigating note fragmentation. For clustering, we construct an affinity matrix by exponentiating cosine similarities between note embeddings, matching the formulation in our clustering loss \eqref{eq:affinityLoss}. Figure~\ref{fig:postprocess} illustrates the results of frame-level and note-level transcription, with the latter nearly perfectly reconstructing compared to the former's fragmented notes.

The last issue remains unresolved, as our attempted iterative matching-filter approach suffers from category merging and is limited to frame-level processing; thus, we do not discuss it further.

\section{Evaluation}

\subsection{Experimental Setup}

\subsubsection{Datasets}
Table~\ref{tab:dataset} summarizes the datasets used in our experiments. We train primarily on MusicNet \citep{MusicNet}, but replace its DTW-aligned annotations, which are known to be inaccurate, with the refined labels from MusicNetEM \citep{MusicNetEM}. All audio is resampled to 22,050 Hz and split into non-overlapping 900-frame, i.e., 10.4-second clips (frame size: 256 samples, matching the CQT hop length). Tracks sharing the same instrument are mixed to form a single timbral class per piece.

For evaluation, we use three real-world polyphonic datasets with solo tracks: BACH10 \citep{bach10} (fixed quartet of violin, bassoon, clarinet, and saxophone), URMP \citep{URMP}, and PHENICX \citep{phenicx}, with the latter featuring complex orchestral mixtures with an average of 9.5 instrument classes per piece.

\begin{figure}[t]
  \centering
  \includegraphics[width=\columnwidth]{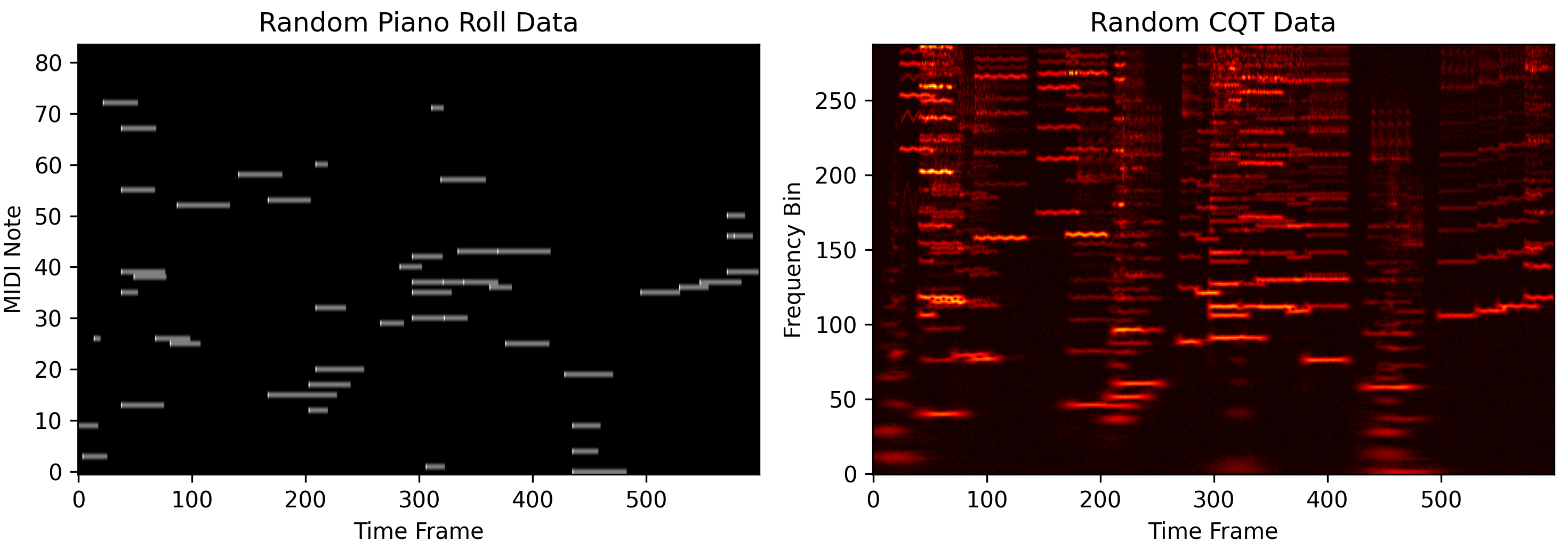}
  \caption{Randomly generated piano-roll (left) and the corresponding CQT spectrogram of the synthesized audio using a trumpet timbre (right).}
\label{fig:dataview}
\end{figure}

Motivated by Slakh2100 \citep{Slakh2100}, a large-scale fully synthesized dataset, we question the necessity of human-composed, real-recorded data. We therefore generate synthetic audio via a simple algorithm that randomly places notes across our 84-note pitch range and adds harmonically overlapping chord tones with probability. MIDI sequences are rendered with \texttt{FluidSynth}, applying randomization in dynamics, pitch tuning, and articulation to improve realism. Each sample uses a single timbre; an example appears in Figure~\ref{fig:dataview}.

For the timbre-agnostic model, we synthesize 252 clips with 900 frames each from 33 General MIDI programs, yielding $\sim$24.1 hours of training data. For timbre-separated transcription, we define 10 categories of acoustically similar instruments, each contributing 12 clips with 600 frames each, totaling $\sim$13.9 minutes of base material. During training, clips are dynamically mixed by blind addition across categories with random gain scaling; all possible combinations are exhaustively enumerated across epochs, greatly increasing data diversity.

\begin{table}[t]
\centering
  \begin{tabular}{ccccc}
  \toprule
  \bfseries Dataset & \bfseries Dur. & \bfseries Songs & \bfseries Instr. & \bfseries K/Song \\ \midrule
  \textbf{MusicNet} & 34h & 330 & 11 & 1–8 \\
  BACH10 & 334s & 10 & 4 & 4 \\
  PHENICX & 637s & 4 & 10 & 8-10 \\
  URMP & 1.3h & 44 & 14 & 2–4 \\
  \textbf{Our:AMT} & 24h & 8316 & 33 & 1 \\
  \textbf{Our:Sep} & 836s & 120 & 34 & 1 \\
  \bottomrule
  \end{tabular}
  \caption{Datasets used in experiments. ``Our:AMT'' and ``Our:Sep'' are our synthetic datasets created to examine whether human composition and real recordings are indispensable. The latter’s timbres are grouped into 10 classes; when training, cross-class mixtures yield single-sample multi-timbre pieces. Bold datasets are used for training.}
\label{tab:dataset}
\end{table}

\begin{table*}[t!]
\centering
\resizebox{\textwidth}{!}{%
    \begin{tabular}{*{11}{c}}
    \toprule
    \bfseries Dataset & \multicolumn{2}{c}{\bfseries BACH10 tutti} & \multicolumn{2}{c}{\bfseries BACH10 stems} & \multicolumn{2}{c}{\bfseries PHENICX} & \multicolumn{2}{c}{\bfseries URMP tutti} & \multicolumn{2}{c}{\bfseries URMP stems} \\ 
    \cmidrule(lr){2-3} \cmidrule(lr){4-5} \cmidrule(lr){6-7} \cmidrule(lr){8-9} \cmidrule(lr){10-11}
    \bfseries Metric(\%) & \bfseries $F_F$ & \bfseries $F_N$  & \bfseries $F_F$ & \bfseries $F_N$ & \bfseries $F_F$ & \bfseries $F_N$ & \bfseries $F_F$ & \bfseries $F_N$ & \bfseries $F_F$ & \bfseries $F_N$ \\ 
    \midrule
    Ours & 84.6 & 75.2 & 91.6 & 88.9 & 63.2 & 49.4 & 75.4 & 71.4 & 81.1 & 83.3 \\
    noLog & 80.7 & 68.1 & 90.0 & 86.6 & 61.1 & 45.7 & 72.4 & 66.7 & 80.7 & 82.1 \\
    BN & 85.9 & 76.4 & 91.9 & 88.6 & 58.5 & 46.0 & 72.6 & 67.8 & 78.5 & 80.3 \\
    Conv39 & 84.1 & 74.6 & 90.8 & 85.8 & 62.1 & 46.2 & 75.4 & 70.6 & 81.6 & 83.0 \\
    BPloss & 84.3 & 63.6 & 91.2 & 77.0 & 44.6 & 46.2 & 74.5 & 61.7 & 81.1 & 71.2 \\
    \rowcolor{red!8}
    MN & 84.8 & 79.4 & 88.4 & 86.7 & 69.7 & 55.6 & 79.7 & 77.4 & 82.3 & 84.4 \\
    \textit{l}CQT & 79.1 & 63.5 & 88.4 & 83.5 & 63.2 & 47.9 & 74.8 & 68.1 & 81.4 & 81.9 \\
    \rowcolor{red!8}
    $l\text{CQT}_\text{MN}$ & \textbf{86.5} & 79.2 & 89.1 & 87.9 & \textbf{70.1} & 56.3 & 79.5 & 75.9 & 81.6 & 82.5 \\
    \cmidrule(lr){1-11}
    BP & 84.8 & 53.7 & 92.0 & 26.9 & 58.0 & 44.9 & 70.7 & 60.4 & 80.9 & 76.8 \\
    $\text{BP}^\text{fl}$ & 86.1 & 75.2 & \textbf{92.3} & 89.3 & 62.2 & 47.7 & 76.2 & 71.5 & 82.2 & 84.9 \\
    \rowcolor{red!8}
    $\text{BP}^\text{fl}_\text{MN}$ & 85.7 & \textbf{80.5} & 89.3 & 87.8 & 69.8 & 54.0 & \textbf{79.9} & \textbf{77.9} & \textbf{82.8} & 84.9 \\
    \cmidrule(lr){1-11}
    OF & 82.7 & 70.7 & 91.4 & 86.6 & 55.9 & 47.6 & 67.9 & 64.4 & 81.1 & 80.7 \\
    $\text{OF}^\text{fl}$ & 82.1 & 72.3 & 91.8 & \textbf{90.0} & 55.5 & 47.0 & 70.1 & 66.7 & 81.2 & 83.7 \\
    \rowcolor{red!8}
    $\text{OF}^\text{fl}_\text{MN}$ & 84.6 & 76.7 & 87.9 & 88.6 & 65.5 & \textbf{56.7} & 76.3 & 74.1 & 81.1 & \textbf{86.1} \\
    \bottomrule
    \end{tabular}%
}
\caption{Comparison of timbre-agnostic transcription. Unless stated otherwise, models are trained on our synthetic dataset. \textbf{``noLog''}: EnergyNorm without log; \textbf{``BN''}: BatchNorm replacing EnergyNorm (\texttt{BasicPitch}-style); \textbf{``Conv39''}: \texttt{BasicPitch}'s 39-tap conv replacing our dilated conv; \textbf{``BPloss''}: \texttt{BasicPitch}'s loss; \textbf{``MN''}: trained on MusicNet; \textbf{``\textit{l}CQT''}: learnable CQT; \textbf{``\textit{l}$\text{CQT}_\text{MN}$''}: learnable CQT trained on MusicNet; \textbf{``BP''}: re-implementation of \texttt{BasicPitch} \citep{BasicPitch}; \textbf{``$\text{BP}^\text{fl}$''}: BP with focal loss; \textbf{``$\text{BP}^\text{fl}_\text{MN}$''}: $\text{BP}^\text{fl}$ trained on MusicNet; \textbf{``OF''}: re-implementation of \texttt{Onsets\&Frames} \citep{OnsetsAndFrames} with its original BCE loss; \textbf{``$\text{OF}^\text{fl}$''}: OF with focal loss; and \textbf{``$\text{OF}^\text{fl}_\text{MN}$''}: $\text{OF}^\text{fl}$ trained on MusicNet. In headers, ``tutti'' = full-mix polyphonic pieces; ``stems'' = single-instrument samples.}
\label{tbl:evaluate_basicamt_tbl}
\end{table*}

\begin{table*}[t]
\centering
\resizebox{\textwidth}{!}{%
    \begin{tabular}{*{16}{c}}
    \toprule
    \bfseries Dataset & \multicolumn{3}{c}{\bfseries BACH10 2mix} & \multicolumn{3}{c}{\bfseries BACH10 3mix} & \multicolumn{3}{c}{\bfseries BACH10 4mix} & \multicolumn{3}{c}{\bfseries URMP 2mix} & \multicolumn{3}{c}{\bfseries URMP 3mix} \\ 
    \cmidrule(lr){2-4} \cmidrule(lr){5-7} \cmidrule(lr){8-10} \cmidrule(lr){11-13} \cmidrule(lr){14-16}
    \bfseries Metric(\%) & \bfseries $F_{FS}$ & \bfseries $F_S$ & \bfseries ratio & \bfseries $F_{FS}$ & \bfseries $F_S$ & \bfseries ratio & \bfseries $F_{FS}$ & \bfseries $F_S$ & \bfseries ratio & \bfseries $F_{FS}$ & \bfseries $F_S$ & \bfseries ratio & \bfseries $F_{FS}$ & \bfseries $F_S$ & \bfseries ratio \\ 
    \midrule
    Ours & \textbf{83.4} & 84.6 & 99.0 & \textbf{77.8} & \textbf{80.1} & \textbf{96.7} & 66.2 & 72.7 & 89.9 & 68.9 & 66.8 & 82.6 & 58.5 & \textbf{60.0} & \textbf{77.1} \\
    D16 & 83.2 & 84.4 & 98.8 & 76.8 & 79.5 & 96.0 & 64.5 & 70.4 & 87.1 & \textbf{69.1} & \textbf{68.4} & \textbf{84.7} & 58.0 & 59.5 & 76.5 \\
    MSE & 80.4 & 82.2 & 96.3 & 70.8 & 75.7 & 91.4 & 59.0 & 67.9 & 84.0 & 65.4 & 65.4 & 80.9 & 53.9 & 57.1 & 73.5 \\
    Syn & 72.4 & 78.1 & 91.4 & 58.7 & 68.0 & 82.1 & 46.0 & 56.1 & 69.4 & 49.6 & 53.9 & 66.7 & 40.9 & 45.1 & 58.1 \\
    Rescale & 83.4 & \textbf{84.6} & \textbf{99.1} & 76.8 & 79.6 & 96.2 & 63.6 & 70.9 & 87.7 & 68.4 & 66.8 & 82.7 & \textbf{58.6} & 59.8 & 76.9 \\
    Share & 82.1 & 83.0 & 98.4 & 76.1 & 78.9 & 96.6 & \textbf{68.5} & \textbf{72.8} & \textbf{91.0} & 69.0 & 68.3 & 84.6 & 57.2 & 57.5 & 74.1 \\
    \cmidrule(lr){1-16}
    Tanaka & 77.9 & 79.6 & 93.2 & 66.1 & 69.0 & 83.4 & 55.4 & 59.2 & 73.2 & 65.5 & 64.1 & 79.3 & 56.5 & 56.5 & 72.8 \\
    \bottomrule
    \end{tabular}%
}
\caption{Comparison of timbre-separated transcription. Unless stated otherwise, models are trained on MusicNet with InfoNCE loss \eqref{eqn:infonceloss}. \textbf{``D16''}: $D=16$; \textbf{``MSE''}: using $L_{\text{affinity}} $\eqref{eq:affinityLoss}; \textbf{``Syn''}: trained on our synthetic dataset; \textbf{``Rescale''}: forcibly scaling amplitude using Frame prediction before InstanceNorm in timbre encoding branch; \textbf{``Share''}: sharing the first residual block between two branches; \textbf{``Tanaka''}: baseline model \citep{DeepSphericalClustering} using InfoNCE. All experiments use identical pre-trained AMT branch parameters except ``Share''.}
\label{tbl:evaluate_septimbre}
\end{table*}

\subsubsection{Evaluation Metrics}
We evaluate transcription performance at frame and note levels using \texttt{mir\_eval} \citep{mir_eval}. Since \texttt{BasicPitch}’s note creation requires thresholds for frames and onsets, we employ a coarse-to-fine search strategy to identify optimal thresholds that maximize model performance, assuming the metric is concave in the threshold. The search converges to $10^{-5}$ precision: first optimizing the frame threshold by maximizing frame-wise $F_1$ ($F_F$), then fixing it while tuning the onset threshold to maximize note-level $F_1$ ($F_N$). A note is correct if its pitch matches ground truth and its onset is within $\pm 50$\,ms of the reference.

For timbre-separated transcription, we extract note events following Section~\ref{sec:postprocess} using the optimized thresholds. The number of clusters $K$ is set manually. To resolve the label permutation problem, we align estimated and reference piano-roll matrices by minimizing MSE over all track permutations. The separated note-level $F_1$ score ($F_S$) is then computed as the average across tracks. Furthermore, to evaluate the effectiveness of our postprocessing, we also compute the note-level $F_1$ score ($F_{FS}$) of notes generated from frame-level clustering results.

Our core metrics are $F_F$, $F_N$, and $F_S$. To isolate the contribution of timbre encoding, we report the ratio $F_S / F_N$. All results are averaged over at least three independent training runs.

\subsection{Implementation Details}
\subsubsection{Parameter Setting}
We use the parameter-efficient downsampling CQT architecture of \citet{CQT2010} with a hop size of 256 samples ($\sim$11 ms at 22,050 Hz). The timbre embedding dimension is set to $D = 12$. Training and validation data follow a 10:1 ratio. We optimize with AdamW with initial $lr = 3 \times 10^{-4}$. Training lasts 60 epochs with a batch size of 18, and the model is selected based on the minimum validation loss. For timbre-separated transcription on our synthetic dataset, we generate mixtures of two or three timbres per sample and add mild white noise. The final model comprises: CQT (19,944 parameters), AMT branch (18,978), and timbre encoder (25,983).

\subsubsection{Timbre-Agnostic Transcription Baseline}
We re-implement \texttt{BasicPitch} \citep{BasicPitch} and \texttt{Onsets\&Frames} \citep{OnsetsAndFrames} in PyTorch as timbre-agnostic baselines.

We remove the pitch prediction neck from \texttt{BasicPitch}, following the original observation that pitch supervision is non-essential, yielding a 56,517-parameter model ($\sim\text{3}\times$ the size of ours). Experiments labeled with ``BP'' use this architecture.

For \texttt{Onsets\&Frames}, we retain only the frame and onset branches and adapt the network to our input configuration, resulting in a 1,714,076-parameter model. Experiments prefixed with ``OF'' employ this architecture.

\subsubsection{Timbre-Separated Transcription Baseline}
We reproduce the method of \citet{DeepSphericalClustering}. Due to the absence of official code and unspecified hyperparameters, we set the BiLSTM hidden size to 256, STFT window length to 1024 (matching the original frequency resolution), hop size to 256 (consistent with our setup), and embedding size to 12. We replace the original loss with our $L_{\text{InfoNCE}}$ and use our AMT branch for multi-pitch estimation. The resulting timbre encoder contains 4,897,776 trainable parameters.

\subsection{Overall Performance Comparison}
Table~\ref{tbl:evaluate_basicamt_tbl} presents the results for timbre-agnostic transcription. Comparing the rows ``Ours'', ``$\text{BP}^\text{fl}$'', and ``$\text{OF}^\text{fl}$'' (all trained with focal loss) and the additional rows ``MN'', ``$\text{BP}^\text{fl}_\text{MN}$'' and ``$\text{OF}^\text{fl}_\text{MN}$'' (trained on MusicNet), we observe that our model achieves performance on par with \texttt{BasicPitch} using significantly fewer parameters. Moreover, our CNN-based architecture substantially outperforms the RNN-based \texttt{Onsets\&Frames}, confirming the efficiency of our design. 

A similar trend appears in timbre-separated transcription in Table~\ref{tbl:evaluate_septimbre}: our method (``Ours'') markedly surpasses the RNN-based baseline of \citet{DeepSphericalClustering} (``Tanaka''). Notably, during training, both RNN baselines exhibit an earlier rise in validation loss, indicating overfitting to the training set. Given their much larger parameter counts, this suggests limited generalization under data constraints and further underscores the robustness, sample efficiency, and trainability of our lightweight architecture.

\subsection{Ablation Study on Core Components}
\paragraph{EnergyNorm}
Comparing rows ``Ours'' and ``noLog'' in Table~\ref{tbl:evaluate_basicamt_tbl}, we find that non-negative features indeed constrain the model’s expressive capacity. We further compare against BatchNorm (row ``BN''), which is used in \texttt{BasicPitch}. While ``BN'' slightly outperforms our method on BACH10, it underperforms significantly on the other two test sets. Inspired by \citet{DomainIncrementalLearning}, we hypothesize that our synthetic data distribution aligns more closely with BACH10, and because BatchNorm stores learned statistics for inference, it suffers from poor generalization to more divergent domains, highlighting the advantage of our normalization strategy in cross-dataset robustness.

\paragraph{Dilated Convolution}
Replacing \texttt{BasicPitch}'s big kernel with our frequency-dilated kernel (rows ``Ours'' and ``Conv39'') yields slightly better performance despite fewer parameters, confirming the benefit of an expanded receptive field. Moreover, non-dilated variants consistently require higher thresholds (not shown), likely because they cannot directly model octave-spanning context, leading to stronger harmonic ghosts (false positive) that must be suppressed post-hoc.

\paragraph{Focal Loss}
While the sophisticated postprocessing in \texttt{BasicPitch} may attenuate the undesirable impact of the blurred onset predictions produced by its original loss (Figure~\ref{fig:loss}), such mitigation is inherently limited. Table~\ref{tbl:evaluate_basicamt_tbl} reveals that models trained with focal loss (``Ours'', ``$\text{BP}^\text{fl}$'', ``$\text{OF}^\text{fl}$'') consistently achieve significantly higher $F_N$ than their counterparts without it, demonstrating that accurate onset estimation remains crucial for high-quality note creation.

\paragraph{InfoNCE Loss}
Comparing rows ``Ours'' and ``MSE'' in Table~\ref{tbl:evaluate_septimbre}, which correspond to $L_{\text{InfoNCE}}$ and $L_{\text{affinity}}$ respectively, shows that $L_{\text{InfoNCE}}$ consistently and significantly outperforms $L_{\text{affinity}}$.

\paragraph{Postprocessing}
As shown in Table~\ref{tbl:evaluate_septimbre}, $F_{FS}$ is almost always lower than $F_{S}$, indicating that our postprocessing method is not only faster but also yields better performance. This is also illustrated in the first row of Figure~\ref{fig:tsne}, where note-level aggregation yields more separable embeddings, significantly improving clustering robustness.

\subsection{Exploratory and Negative Results Analysis}
\subsubsection{Limitations of Synthetic Data}
Synthetic data underperforms MusicNet due to two domain gaps. First, our synthetic notes span the full 84-note range, whereas real instruments occupy limited registers; timbre is approximately consistent only within a narrow pitch range \citep{invariantHarmonic}, leading to fragmented clusters when scattered across octaves. Second, synthesizers produce static timbres lacking the dynamic variation of real performances. These limitations underscore the need for human-composed and real-recorded datasets capturing authentic timbral complexity.

\subsubsection{Learnable CQT}
Motivated by learnable time–frequency representations, we explore whether a learnable CQT can improve performance. To ensure stability, we first train the full network with fixed CQT parameters, then unfreeze them for joint fine-tuning. While training loss decreases noticeably, test performance degrades on our synthetic data (``Ours'' vs. ``\textit{l}CQT''), and yields no improvement on MusicNet (``MN'' vs. ``$l\text{CQT}_\text{MN}$'').

We attribute this to overfitting: the added flexibility enhances model capacity, but with limited training data, it leads to memorization rather than generalization. For lightweight systems, handcrafted CQT parameters thus offer better robustness. The potential benefits of learnable CQT may only emerge in large-scale settings, which remains an open question.

\subsubsection{Exploring Alternative Encoding Architectures}
We evaluate several design choices for the timbre encoding branch; results are summarized in Table~\ref{tbl:evaluate_septimbre}.

\textit{Encoding dimension.} Increasing the embedding dimension to 16 (row ``D16'') yields no significant gain. Notably, \citet{ZeroShot} achieves effective separation with only six dimensions, suggesting that even lighter encodings may suffice.

\textit{Multitask coupling.} Prior work \citep{Cerberus,TimbreTrap} suggests that multitask learning can be mutually beneficial. We test two coupling strategies: 1) sharing the first residual block between branches (``Share''), and 2) rescaling the timbre embeddings using corresponding frame predictions before InstanceNorm (``Rescale''). Neither improves performance and often degrades it, indicating that such coupling may not suit lightweight networks.

\begin{figure}[t]
  \centering
  \includegraphics[width=\columnwidth]{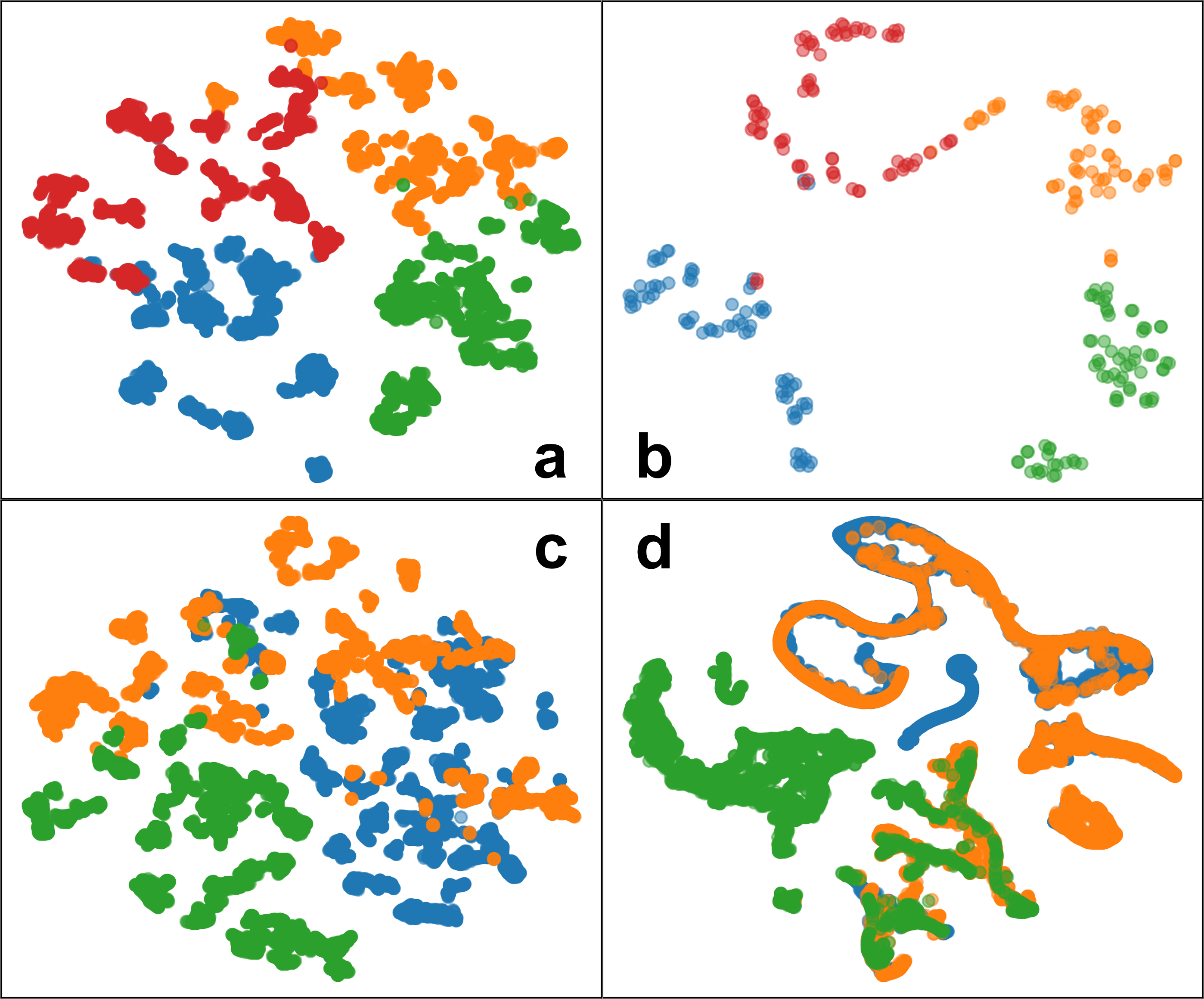}
  \caption{t-SNE visualization of timbre embeddings. (a) Frame-level embeddings for BACH10 Piece 2; (b) note-level aggregates of (a); (c) frame-level for URMP Piece 18; (d) frame-level for URMP Piece 18 using top-$k$ attention.}
\label{fig:tsne}
\end{figure}

\textit{Attention mechanisms.} In light of the contributions of Transformers to timbre modeling \citep{TransformerAgnostic, attentionAMT}, we explore attention \citep{attention} for its potential to incorporate global context and enhance cluster compactness (not included in the table). Given the large $N \times T$ size, we restrict ourselves to linear-complexity variants. All tested configurations significantly hurt performance. Figure~\ref{fig:tsne}(d) shows that self-attention densifies intra-class connections but spuriously links distinct timbre manifolds. This occurs when attention weights connect ambiguous boundary regions where embeddings from different instruments appear similar, e.g., at overlapped notes, thereby bridging otherwise separable clusters. This is particularly detrimental to spectral clustering, which relies on graph connectivity rather than metric separability. To prevent such fusion, we also evaluate a Transformer variant with residual connections on the attention output, which merely maintains baseline performance. Attention might be more effective after note-level aggregation, but that prevents end-to-end training. For lightweight networks, especially when frame-level embeddings are not well separated, such operations appear unnecessary and potentially detrimental to separability.

\subsection{Efficiency}
Our model is simple and lightweight enough to run directly in a web browser. On an Intel\textsuperscript{\textregistered} Ultra 7 255H CPU using Microsoft Edge, timbre-agnostic transcription of a 301-second duet takes 17.1 s, compared to 39.1 s for \texttt{BasicPitch}. The postprocessing step requires 78 ms. For timbre-separated transcription, our model inference takes 38.8 s, followed by 704 ms for clustering-based postprocessing. In contrast, RNN-based baselines suffer from ONNX export issues and have prohibitively large parameter counts, rendering them impractical for real-world deployment.

\section{Reproducibility}
The data and code used in this paper are accessible via \url{https://github.com/madderscientist/timbreAMT} and DOI: \url{https://doi.org/10.5281/zenodo.19229826}.

\section{Conclusion}
We have introduced a compact and efficient architecture for timbre-separated music transcription that overcomes major limitations of current approaches: fixed instrument vocabularies, poor generalization to unseen timbres, and high computational cost. By performing clustering on coherent note events rather than raw time–frequency bins, our method reduces fragmentation, improves separation quality, and supports flexible inference without waveform reconstruction.

Several promising directions remain for future work. First, the current reliance on empirically tuned thresholds for note creation could be replaced by learned, content-adaptive thresholds. Second, training data could be enhanced by mixing instruments in complementary frequency bands to better reflect natural orchestration. Finally, our model still requires the number of instrument classes to be specified at inference time and is unable to handle overlapping situations. The matching-filter approach described at the end of Section~\ref{sec:postprocess} remains a promising direction worth further exploration.

\IfFileExists{\jobname.ent}{
   \theendnotes
}{
}


\small
\bibliography{main}

\end{document}